# Considerations on Resource Usage in Exceptions and Failures in Workflows

**Assist.Prof. Alexandra Fortiş, Ph.D.Candidate**
„Tibiscus" University of Timişoara, România,
**Prof. Alexandru Cicortaş, Ph.D.**
„Vasile Goldiş" Western University of Arad, România,
**Assoc.Prof. Victoria Iordan, Ph.D.**
The West University of Timişoara, România

REZUMAT. Lucrarea prezintă o descriere a unor puncte de vedere aparţinând mai multor autori referitor la eşecurile şi excepţiile care apar în workflow-uri, ca o consecinţă directă a indisponibilităţii resurselor implicate în workflow. Fiecare dintre aceste interpretări este caracteristică unei anumite situaţii, în funcţie de interpretarea pe care o dau autorii eşecurilor şi excepţiilor în workflow-uri care modelează sisteme reale.

## 1 Introduction

When modeling real, dynamic and complex systems the natural approach is to use a workflows-based model. Workflow Management Systems are responsible with the coordination of activities involved in, the data within the framework of the process as well as the resources used.

In a generic framework, resources can be considered as actors, executants or participants to a process. The consumption of a resource during the execution of the process may block the workflow.

In this paper we are going to explore several of the possible solutions that can be adopted in case of workflow failures determined by resource

69

disappearing before the completion of the task they were assigned to, during process execution.

## 2 Managing the Resources for Workflows

Generally, a workflow is designed in isolation, without taking into account that it works in the frame of an organization or it has inter-organizational characteristics.

By analyzing the capabilities of existing commercial business process modeling languages to handle exceptions and failures of a workflow, the general conclusion was that none of the offerings can manage a situation related to resources unavailability. The gap is a direct result of the lack of research in this particular area. The workflow engine and the Workflow Management System are the components that should take care of the resource availability.

The core components of workflow engines in the case of many standards and existent WfMs do not take into account the problem for resource availability and do not dispose for an adequate tool for the resource management.

The exceptions and failures that appear during the workflow execution have to be treated in the real context of the modeled system. Some of the specific problems that appear include the following facts:
- An activity which failed is executed in the general context of the workflow, affecting the whole process. In that case, the results of the previous activities must be preserved and analyzed, when needed;
- A workflow has to be dynamically redesigned, in a new context, by taking into account the resources involved in, which change their availability during the execution of different tasks;
- The human resource has to interact with the workflow due to the failure and manage the information revealed by failures and exceptions, in order to improve the existent workflow.

Due to the fact that the ERP (Enterprise Resource Planning) also manages the resources that usually are in some way different to those used in workflow execution, one of the major problems is that to conceive a model that allows to manage the resources in the general (and real) context of the organization in order to fulfill the requirements of the system, ERP and WfMs.

By resources, the literature in this field refers to both human and non human resources. As presented in [RHEA04], the human resources can be affected by their position within the organization, their privileges, by the affiliation to an organizational team or group or by the relations with other human resources. In the context, the workflow may also fail because of the capabilities, the qualifications and the skills at individual level of the resource.

As for the non human resources, they can be durable and their utilization during the completion of different tasks does not affect the workflow by producing failures. On the other hand, in the case of consumable non human resources, their use may cause the not completion of a task/work item, producing a failure of the workflow.

Using appropriate services that allow the WfMS and the core components of the workflow engine to search the needed resources for activities in execution could be a solution.

Conception for a model that manages all resources from a company is a difficult task that can be considered as an idealistic one, but it is a real necessity. As difficulties care the followings:
- Different measure units for representation and usage;
- Different time scales used;
- Different concepts regarding the capabilities of the resource availability;
- Generation of redundant specifications for resources.

## 3 Exceptions and Failures in Workflows

The workflows constitute a necessary modeling an analysis tool in every complex system. Their complexity implies, in many cases the failures of the systems and also of the workflows. Designing the complex systems and the workflow also must be done in a strict correlation. The automatic recovery in complex systems and workflows does not constitute a real and efficient solution in many cases. The human intervention embraces a lot of aspects that depend on the system and workflow peculiarities [MA03].

Workflow exceptions and workflow failures are situations not predicted in the workflow model or aspects that appear if there is a deviation from the model and the real world. Exceptions represent a deviation from the normal execution of a workflow model, an anomaly occurring in the normal execution of the flow.

For an efficient action in the workflow, the human intervention must be supported with quality information concerning the concrete situation, the status of the workflow engine and some possible recovery solutions.

An exception handler has to solve particular events by analyzing the unexpected situation with respect to a set of situations/similarities that have already been investigated. Despite the diversity of these unexpected events, a set of patterns has been developed 0, in order to investigate and solve the exceptions and failures that are affecting the workflows, at a general level.

In Workflow Management Systems (WfMS) the exception handling, as one of the most important problems, requires:
- a perspective over exceptions and exception handling, taking into account the three organizational levels: strategic, tactic and operational;
- human roles in recovery such as design changes, ad hoc execution and engine status manipulation

and WfMS features concerning:

- Event handlers;
- Situation awareness;
- Problem characterization;
- Recovery toolkit.

From the start, in the earlier approaches [EL95], [EL96], [EL98] the failures and exceptions of workflow were considered as being:

- Basic failures with direct reference to the failures generated by the systems that support the WfMS: operating system, DBMS, network;
- Application failures when the applications that are invoked to execute tasks fail during their execution;
- Expected exceptions [AHEA05] which are events that can be predicted during the modeling phase and that do not correspond to the normal behavior of the process;
- Unexpected exceptions that are generated be the impossibility treating certain situations that the system model does not support it. These types of exceptions cannot be predicted at the modeling stage and may require in some way human intervention.

Exception Handling, as it is recognized by most literature, with the major function of predicting any possible cause of failure or exception during design is very difficult or even impossible task and makes the system very complex and hard to manage [EL98], [Cas98], [KD00]. Therefore, being prepared to deal with failures and exceptions at execution time is a critical factor for WfMS success.

Another analysis, [Chi00] treats the exceptions as being the exception source. It can either be internal or external sources to the workflow. External sources are consequences of failures in components that participate in the WfMS such as operating systems, database management systems, software applications, machines and equipment, or operations in external organizations. Internal sources are directly related to workflow management issues, e.g., being unable to find a component to execute a task, or a missing a deadline.

## 4  Expected and Unexpected Exceptions

Expected exceptions are cases that can be predicted during the modeling stage but do not correspond to the normal process behavior. Some mechanisms should be implemented to handle these situations because they can occur frequently.

In [Cas98] and [CCPP99] the authors have identified few classes of expected exceptions, according to the events that generate them:
- Workflow exceptions generated by the beginning or the finalization of a task;
- Data exceptions generated when workflow relevant data is changed;
- Temporal exceptions generated by the occurrence of a given timestamp;
- External exceptions activated by external events.

An important problem is to identify whether the exception is synchronous or asynchronous to the evolution of the process. The synchronous class of exceptions is the workflow type since it happens after a change in the process state. From our point of view, a synchronous exception appears after the completion of an unfit task. The exhaustion of one or more resources may lead to an incorrect termination of one process and this error can propagate and will affect the entire workflow. Asynchronous exceptions are difficult to model because they are produced by external influences on the workflow Furthermore, such exceptions can arise at any execution moment.

Unexpected exceptions result from inconsistencies between process modeling in the workflow and the actual execution [EL98]. These are also the result from incomplete or design errors, improvements or changes in the business manufacture or quality and customer satisfaction issues not known during the modeling stage. This type of exceptions is frequent in highly complex or dynamic environments. Usually, unexpected exceptions force the process to change to a halt state and require human intervention. In situations where this kind of exceptions occurs frequently, one should consider redesigning the workflow model, if it is out of date, or adopting different technologies based on collaborative work or adaptive workflow systems [CCPP99].

Concerning the organizational perspectives the exceptions can be treated using LDAP directories of the organization. In such a manner the exceptions stated by the earlier approaches [Saa95] can be solved. The major problem here is that to have a good design and an appropriate data base for the information that is used and processed concerning the workflow and the resources. The human interaction with the workflow is required and it represents one of the challenges for workflow designers.

An alternative solution that was proposed is the adoption of dynamic and adaptive workflow systems to react to exceptions during workflow execution. Encountering an exception, the operator have to be able to change the system by either creating a new path for the exceptional process or change all the processes running on the system to the newly created path. A major feature is to keep the systems consistency and correctness during/after the change.

Some other studies have proposed the adoption of dynamic and adaptive workflow systems to react to exceptions during workflow execution as will be shown as follows. In [GAP97] the authors proposed an integrated architecture of formal coordinated processes with informal cooperative processes. When the system identifies that it is not able to proceed with process execution, it gathers all the information concerning the specific situation and generates a flow interrupt. The interrupt triggers a tool able to select the best suitable cooperative technique to handle the specific case. In [Saa95] is given an approach focused on organizational semantics.

In [AHEA05] a different notion was proposed: the worklets as being an "extensible repertoire of self-contained sub-processes and associated selection and exception handling rules, and grounded in a formal set of work practice principles of Activity Theory, to support the modeling, analysis and enactment of business processes".

The notion was developed after pointing out some descriptive and relevant principles, capable to describe the type of participation in organizational work practices.

For the activities involved in the workflow, one has to consider their hierarchical, communal, contextual, dynamic and mediated character. The tasks have to be considered only in the context of the workflow. From the author's point of view, when elaborating the work plan, the main characteristic that has to be implemented is its flexibility within the context. Following this requirement, the work plan associated to a real system can be redesigned each time an exception or a failure is affecting the workflow model. Encountering exception and failures during the execution may lead to the creation of some sort of data base containing information on the situation that has not been yet analyzed. By including this information, the existing workflow model can be improved, exception and failures being eliminated.

The result of implementing those principles is a flexible workflow support system which is sketching the model, but allows further changes. Each time a task is executed, the new concept offers a dynamic list of possible choices, context related, this approach being able to incorporate the exceptions in the model. In this case, exceptions and failures are treated as normal events in the process' evolution and the most appropriate exception handling method can be used for a normal workflow course.

In this approach, each task of a process instance is connected to a flexible and open list of possible actions to be done, depending on the context. In fact, a workflow can be visualized as a collection of interrelated worklets. A worklet is conceived as a small workflow modeling only one of the multiple tasks included in the general workflow process. Through worklets, any task can be created, assigned and reviewed in order to assure the well function of the workflow model.

Each time an exception appears in the execution of the process, a complementary worklet for handling the event can be defined and it can be included dynamically into a running workflow instance on an as-needed basis. Such worklets are created in the same manner as for any normal event in the process, creating a sort of library containing information on exceptions appearing during the execution of the process together with the method used to solve the failures.

More recently in [AHEA07], a dynamic and extensible exception handling for workflows, a service-oriented implementation that is used in Yawl, was proposed.

## 5 Conclusions and Future Work

In our previous studies and articles, a particular case was studied, i.e. the technological flow for a certain food additive (baking powder) industrial production. The major aspect to solve through our past and future research is to offer a novel approach for material and energy balances that are presenting a high level of significance in several industries [For06]. As stated in [For07], the first aspect presented was a Petri net-based model, used to identify the basic properties for the underlying workflow system: safeness, boundedness conservation and liveness.

The most appropriate modeling technique to be used is BPMN, because of its conformance with the workflow patterns, its extensibility, the error handling and its compensation support [FF06].

The next stage we take into account is to extend the analysis from the Petri net-based model to a new one, based on workflow nets. Through this new approach, the meaning of a transition from Petri nets is extended by using four special types of transitions in workflow nets: And-split, And-join, XOR-split and XOR-join, [For07], [CIFF06].

In brief, the models proposed are easily understandable, present a high level of visualization, have the capability to solve complex problems with a large amount of data and restrictions and can be based upon computer systems. But all this aspects are related to an ideal, generic case. We have to offer a solution that will include workflow manipulation for the case when one ore more resources are used/unavailable - aspects that are not entirely covered by the workflow theory, as stated before – or to handle the system in case of different production capabilities.

As for the future work, the first aspect to solve will be to include in the workflow model, via messages, quantitative information related to the amount of resources involved in the process, together with solution in case of resource consumption, that being one of the major causes of a workflow failure.

As was previously stated, many works and projects give solutions depending on the features of the exceptions and their treatment which are convenient to different points of view. As was seen the problem of the failures and exceptions must use information that in time becomes unrealistic due to the evolution of the system. Based on it the future models must treat in a suitable manner this problem. For that, the model needs the tools that allow the workflow engine on the one side and the ERP on the other side to access the enterprise resources.

## References

[AHEA05] **Adams, M., ter Hofstede, A.H.M., Edmond, D., van der Aalst. W.M.P. -** *Facilitating flexibility and dynamic exception handling in workflows through worklets*. In Orlando Bello, Johann Eder, Oscar Pastor, and Joao Falcao e Cunha, editors, Proceedings of the CAiSE'05 Forum, pages 45-50, Porto, Portugal, June 2005.

[AHEA07] **Adams, M., ter Hofstede, A.H.M., Edmond, D., and van der Aalst, W.M.P. -** *Dynamic and Extensible Exception Handling for Workflows: A Service-Oriented Implementation.* (PDF, 311Kb) BPM Center Report BPM-07-03, BPMcenter.org, 2007.

[Chi00] **Chiu, D. K. -** *Exception Handling in an Object-Oriented Workflow Management System*, PhD Thesis, Hong Kong University of Science and Technology, 2000.

[Cas98]    **Casati,F. -** *Models, Semantics, and Formal Methods for the Design of Workflows and Their Exceptions*, PhD Thesis, Politecnico di Milano, 1998.

[CCPP99]   **Casati, F., Ceri, S., Paraboschi, S. and Pozzi, G.** - *Specification and Implementation of Exceptions in Workflow Management Systems*, ACM Transactions on Database Systems, 24, 3, (1999).

[CIFF06]   **Cicortaş, A., Iordan, V., Fortiş, A., Fortiş, F.,** - *Reengineering the Failed Workflows,* International Workshop in Collaborative Systems, Annals of the Tiberiu Popoviciu Seminar, Supplement international Workshop in Collaborative Systems, Vol. 4, Cluj-Napoca,Romania, 2006

[EL95]     **Eder, J., Liebhart, W.** - *The Workflow Activity Model WAMO*, Int. Conf. on Cooperative Information Systems, Vienna, Austria, 1995.

[EL96]     **Eder, J., Liebhart, W. -** *Workflow Recovery,* 1st IFCIS Intl. Conf. on Cooperative Information Systems (CoopIS'96). Brussels, Belgium: IEEE International, 1996.

[EL98]     **Eder, J., Liebhart, W.** - *Contributions to Exception Handler in Workflow Management,* Int. Conf. on Extended Database Technology (EDBT'98), Workshop on Workflow Management Systems, Valencia, Spain, 1998.

[FF06]     **Fortiş, A., Fortiş, F.,** - *Using BPM Technologies for Material Balances Modelling,* Proceedings of the 8[th] International Symposium on Symbolic and Numeric Algorithms for Scientific Computing (SYNASC 2006), Timişoara. IEEE Computer Society 2006,

[For06]    **Fortiş, A., -** *Business Process Modeling Notation – An Overview*, Proceedings of the 1[st] European Conference on Computer Science and Applications, XA2006, Timişoara, România, 2006.

[For07]   **Fortiş, A.,** - *Analysis of some properties for a basic Petri net model*, Analele Universitatii Tibiscus, Seria Informatica, 2007, Timisoara, Romania

[GAP97]   **Guimaraes, N., Antunes, P., and Pereira, A. P. -** *The Integration of Workflow Systems and Collaboration Tools*, in A. K. Dogac, Leonid Ozsu (Eds.), Advances in Workflow Management Systems and Interoperability, Istambul, 1997.

[KD00]    **Klein, M., and Dellarocas, C.** - *A Knowledge-Based Approach to Handling Exceptions in Workflow Systems*, Computer Supported Cooperative Work, 9, 3 (2000).

[MA03]    **Mourao, H., Antunes, P.** - *Workflow Recovery Framework for Exception Handling: Involving the User, in Groupware: Design, Implementation and Use*, Lecture Notes in Computer Science, 2003.

[RHEA04]  **Russel, N., ter Hofstede, A.H.M., Edmond, D., van der Aalst, W.M.P. -** *Workflow Resource Patterns*, BETA Working Paper Series, WP 127, Eindhoven University of Technology, Eindhoven, 2004.

[RAH06]   **Russell, N., van der Aalst, W.M.P., ter Hofstede, A.H.M. -** *Exception Handling Patterns in Process-Aware Information Systems*, BPM Center Report BPM-06-04, BPMcenter.org, 2006.

[Saa95]   **Saastamoinen, H. -** *On the Handling of Exceptions in Information Systems*, PhD Thesis, University of Jyvaskyla, 1995.